%% file: no_comments.tex
\documentclass[twocolumn,10pt,letterpaper]{article}
\usepackage[top=2cm, bottom=2cm, left=1.8cm, right=1.8cm]{geometry}
\usepackage{graphicx}
\usepackage{float}
\usepackage[keeplastbox]{flushend}
\usepackage{paralist}
\usepackage{times}
\usepackage[hyphens]{url}
\usepackage{indentfirst}
\usepackage{amsmath}
\usepackage{amsfonts}
\usepackage{multirow}
\usepackage{xcolor}
\usepackage{xpatch}
\usepackage{array,multirow,graphicx,adjustbox}
\usepackage{booktabs,fixltx2e}
\usepackage[bf]{caption}
\usepackage[flushleft]{threeparttable}
\usepackage{xcolor}
\usepackage{csquotes} 
\usepackage{pdflscape}
\usepackage{tabularx}
\usepackage{tcolorbox}
\usepackage{graphicx}
\usepackage{tabu}
\usepackage{etoolbox}
\usepackage{url}
\usepackage{subfigure}

\usepackage[marginal,hang,stable]{footmisc}
\renewcommand{\footnoterule}{%
  \kern -3pt
  \hrule width 1in
  \kern 2pt
}

\makeatletter
\def\url@leostyle{%
  \@ifundefined{selectfont}{\def\UrlFont{}}%
  {\def\UrlFont{}}%
}
\makeatother
\usepackage{titlesec}
\titlespacing*{\section}{0pt}{*2}{5pt} 
\titlespacing{\subsection}{0pt}{*2}{4pt}
\titlespacing{\subsubsection}{0pt}{*1.5}{3pt}

\newcommand{\descr}[1]{\medskip\noindent\textbf{#1}}
\newcommand{\descrit}[1]{\smallskip\noindent\textbf{\em #1}}

\newcommand{\reduceA}{\vspace*{-0.00cm}}

\let\OLDthebibliography\thebibliography
\renewcommand\thebibliography[1]{
  \OLDthebibliography{#1}
  \setlength{\parskip}{0.5pt}
  \setlength{\itemsep}{0pt plus 0.3ex}
}

\usepackage{etoolbox} 
\makeatletter
\patchcmd\maketitle{\@makefntext}{\@@@ddt}{}{}
\patchcmd\maketitle{\rlap}{\mbox}{}{}
\makeatother

\definecolor{darkred}{RGB}{123,0,20}
\definecolor{darkblue}{RGB}{0,0,149}
\usepackage[hidelinks,linkcolor = darkblue]{hyperref}
\newcommand{\mycite}[1]{\href{}{\color{darkred}{\cite{#1}}}}%

\begin{document}

\sloppy 

\title{\bf Systematizing Genome Privacy Research:\\A Privacy-Enhancing Technologies Perspective\thanks{Published in the Proceedings on Privacy Enhancing Technologies (PoPETs), Vol. 2019, Issue 1.}}
\date{}
\author{Alexandros Mittos$^1$, Bradley Malin$^2$, Emiliano De Cristofaro$^1$\\[1ex]
\normalsize $^1$University College London~~~$^2$Vanderbilt University}

 \maketitle

\begin{abstract}
Rapid advances in human genomics are enabling researchers to gain a better understanding of the role of the genome in our health and well-being, stimulating hope for more effective and cost efficient healthcare. However, this also prompts a number of security and privacy concerns stemming from the distinctive characteristics of genomic data. 
To address them, a new research community has emerged and produced a large number of publications and initiatives.
In this paper, we rely on a structured methodology to contextualize and provide a critical analysis of the current knowledge on privacy-enhancing technologies used for testing, storing, and sharing genomic data, using a representative sample of the work published in the past decade.
We identify and discuss limitations, technical challenges, and issues faced by the community, focusing in particular on those that are inherently tied to the nature of the problem and are harder for the community alone to address.
Finally, we report on the importance and difficulty of the identified challenges based on an online survey of genome data privacy experts.
\end{abstract}

\reduceA\reduceA\reduceA\reduceA\reduceA\reduceA\reduceA\reduceA\reduceA
\section{Introduction}\label{sec:introduction}
\reduceA\reduceA\reduceA

Facilitated by rapidly dropping costs, genomics researchers have made tremendous progress over the past few years toward mapping and studying the human genome.
Today, the long-anticipated ``genomic revolution''~\cite{wired2} is taking shape in a number of different contexts, ranging from clinical and research settings to public and private initiatives. %

At the same time, the very same progress also prompts important privacy, security, and ethical concerns. 
Genomic data is hard to anonymize~\cite{gymrek2013identifying,shringarpure2015privacy} and contains information related to a variety of factors, including ethnic heritage, disease predispositions, and other phenotypic traits~\cite{fowler}.
Moreover, consequences of genomic data disclosure are neither limited in time nor to a single individual; due to its hereditary nature, an adversary obtaining a victim's genomic data can also infer a wide range of features that are relevant to her close relatives as well as her descendants. 
As an artifact, disclosing the genomic data of a single individual will also put the privacy of others at risk~\cite{golden}. %

Motivated by the need to reconcile privacy with progress in genomics, researchers have initiated investigations into solutions for securely testing and studying the human genome.
Over the past few years, the genome privacy community has produced a relatively large number of publications on the topic, with several dedicated events, e.g., international seminars~\cite{dag1,dag2}, a competition series, 
and the GenoPri workshop now entering its fifth year.\footnote{See \url{https://idash.ucsd.edu/} and \url{https://genopri.org}.}

At the same time, the community is partially operating ahead of the curve, proposing the use of privacy-enhancing technologies (PETs) in envisioned, rather than existing, settings.
In fact, as discussed in this paper, genome privacy research also makes assumptions for the future, e.g., that cheap, error-free whole genome sequencing will soon be available to private citizens, or that individuals will be sequenced at birth so that all genetic tests can be easily and cheaply done via computer algorithms. %

Based on these developments, it is time to take stock of the state of the field.  
To do so, we conduct a systematic analysis of genome privacy research, aiming to evaluate not only what it has achieved so far, but also future directions and the inherent challenges the field faces.
Overall, our work is driven by three main research objectives:
\begin{enumerate}
\item Critically review, evaluate, and contextualize genome privacy research using a structured methodology that can be reused in the future to assess progress in the field.
\item Reason about the relevance of the proposed solutions to current public sequencing initiatives as well as the private market.
\item Identify limitations, technical challenges, and open problems faced by the community. In particular, we aim to assess which of these are likely to be addressed via natural progress and research follow-ups and which are inherently tied to the very nature of the problem, involving challenging tradeoffs and roadblocks.
\end{enumerate}

\descr{Roadmap.} With these objectives in mind, we set out to critically evaluate work produced by the genome privacy community across several axes, using a set of systematic criteria that span a broad spectrum of properties. 
Rather than presenting an exhaustive review of the very large body of work in this field, we adopt a methodology to analyze themes of genome privacy research using a sample of representative papers. 
We focus on research relying on PETs in the context of testing, storing, and sharing genomic data. 
To do so, we retrieve the list of publications in the field from a community managed website (\url{GenomePrivacy.org}) while intentionally excluding papers about attacks and risk quantification (Section~\ref{sec:methodology}).
After identifying relevant sub-areas of genome privacy research, we select results that provide a meaningful sample of the community's work in each area (Section~\ref{sec:stateoftheart}).

Next, we present a systematization which we rely upon to summarize the critical analysis and guide the examination of 10 key aspects of genome privacy (Section~\ref{sec:analysis}).
Finally, in Section~\ref{sec:directions}, aiming to validate and broaden the discussion around the identified challenges, we report on an online-administered survey of genome privacy experts, whom we ask to weigh in on them with respect to their importance and difficulty.
Overall, our analysis, along with the input from the experts, motivates the need for future work as well as more interdisciplinary collaborations,  
while pointing to specific technical challenges and open problems; moreover, our methodology can be reused to revisit new results and assess the progress in the field.

\descr{Main Findings.} Our analysis also helps us draw some important conclusions. 
First, that the effective use of PETs in the context of genome privacy is often hindered by the obstacles related to the unique properties of the human genome. 
For instance, the sensitivity of genomic data does not degrade over time, thus prompting serious challenges related to the lack of effective long-term security protection, as available cryptographic tools are not suitable for this goal. 
Second, we find that the majority of the proposed solutions, aiming to scale up to large genomic datasets, need to opt for weaker security guarantees or weaker models.
While it is not unreasonable to expect progress from the community with respect to underlying primitives, it is inherently hard to address the limitations in terms of utility and/or flexibility on the actual functionalities. 
When combined with assumptions made about the format and the representation of the data, this poses major hurdles against real-life adoption.

On the positive side, we highlight how, in its short lifespan, the genome privacy community has achieved admirable progress.
For instance, several tools can already enable genomics-related applications that are hard or impossible to support because of legal or policy restrictions.

\reduceA\reduceA\reduceA\reduceA\reduceA\reduceA\reduceA\reduceA
\section{Background}\label{sec:background}
\reduceA\reduceA\reduceA\reduceA

\smallskip\noindent\textbf{Progress in Genomics.} In clinical settings, progress in genomics has allowed researchers to link mutations to predisposition to various forms of diseases, including cancer~\cite{de2010predicting}, as well as to response to certain treatments~\cite{zeuzem2014simeprevir}.
These developments provide support for a new \textit{precision medicine} era, where diagnosis and treatment can be tailored to individuals based on their genome, and thus enable more cost efficient, as well as effective, healthcare~\cite{ashley2016towards}.
Moreover, with the costs of whole genome sequencing (i.e., determining the whole DNA sequence of an individual) now in the order of \$1,000~\cite{sequecing2017costs}, clinicians have a greater opportunity to diagnose and treat patients affected by rare genetic disorders~\cite{bick2017successful,gilissen2014genome}.

\descr{Sequencing Initiatives.} The promise of improved healthcare has also encouraged ambitious sequencing initiatives, aiming to build biorepositories for research purposes.
In 2015, the US government announced the Precision Medicine Initiative (now known as the All Of Us Research Program~\cite{allofus2017}), aiming to collect health and genetic data from one million citizens.
Similarly, Genomics England is sequencing the genomes of one hundred thousand patients, focusing on rare diseases and cancer~\cite{england2015100}.  
The rise of data-driven genomics research also prompts the need to facilitate data sharing. 
In 2013, the Global Alliance for Genomics and Health (GA4GH) was established with an objective to make data sharing between institutes simple and effective~\cite{ga4gh}.
The GA4GH has developed various software, such as Beacon\footnote{\url{https://beacon-network.org/}}, which permits users to search if a certain allele exists in a database hosted at a certain organization, and the Matchmaker Exchange~\cite{philippakis2015matchmaker}, which facilitates rare disease discovery.

\descr{Private Market.} Progress has also encouraged the rise of a flourishing private sector market.
Several companies operate successfully in the business of sequencing machines (e.g., Illumina), genomic data storage and processing (e.g., Google Genomics), or AI-powered diagnostics (e.g., Sophia Genetics).
At the same time, others offer genetic testing {\em directly} to their customers, without involving doctors or genetics experts in the process. 
There are now hundreds of direct-to-consumer (DTC) genetic testing companies~\cite{isogg2017}, with prominent examples including 23andMe and AncestryDNA, which have amassed several million customers.\footnote{See \url{http://ancstry.me/2iD4ITy} and \url{https://goo.gl/42Lz9v}.}

\descr{Legal Aspects.} 
The new General Data Protection Regulation (GDPR)~\cite{GDPR2016} has come into effect in the EU in May 2018. 
Its impacts on genetic testing and genomic research are not yet clear. 
The {\em data minimization} principle suggests that the minimum required amount of data should be stored to achieve the intended goal, while the \emph{purpose limitation} principe dictates that researchers should predetermine the scope of the study (Article 5). 
Under Article 35, genetic data is designated both as sensitive and personal data.
As is the case in general with GDPR, we do not yet know how it will affect operational aspects of systems and products, although genomic privacy research could and should assist this process.
In the US, there is no equivalent of GDPR, however, certain legislation and policy protects the privacy of study participants using indirect means. 
For example, to access sensitive data in NIH databases, a researcher must first submit a request. 
Moreover, the Genetic Information and Nondiscrimination Act of 2008 (GINA) states that it is illegal for employers or health insurers to request genetic information of individuals or of their family members. However, this legislation does not cover the cases of life, disability, and long-term care insurance.

\descr{Attacks against Genome Privacy.} 
A few {\em re-identification} attacks have been proposed whereby an adversary recovers the identity of a target by relying on quasi-identifiers, such as demographic information (e.g., linking to public records such as voter registries), data communicated via social media, and/or search engine records~\cite{sweeney2013identifying}. 
For instance, Gymrek et al.~\cite{gymrek2013identifying} infer the surnames of individuals from (public) anonymized genomic datasets by profiling short tandem repeats on the Y chromosome while querying genealogy databases.
Also, in {\em membership inference} attacks, an adversary infers whether a targeted individual is part of a study that is possibly associated with a disease, even from aggregate statistics.
Homer et al.~\cite{homer2008resolving} do so by comparing the target's profile against the aggregates of a study and those of a reference population obtained from public sources. 
Wang et al.~\cite{wang2009learning} leverage correlation statistics of a few hundreds SNPs, while Im et al.~\cite{im2012sharing} use regression coefficients.
Shringarpure and Bustamante~\cite{shringarpure2015privacy} present inference attacks against Beacon by repeatedly submitting queries for variants present in the genome of the target, whereas, Backes et al.~\cite{backes2016membership} attacks focused on microRNA expressions.
More generally, Dwork et al.~\cite{dwork2015robust} prove that membership attacks can be successful even if aggregate statistics are released with significant noise. 
For a comprehensive review of the possible/plausible attacks against genome privacy, we refer the readers to~\cite{erlich2014routes}.

\descr{Genomics Primer.} To assist the reader, Appendix~\ref{primer} provides a primer on terms and concepts commonly used in genomics. %

\reduceA\reduceA\reduceA\reduceA\reduceA\reduceA\reduceA\reduceA\reduceA
\section{Methodology}\label{sec:methodology}
\reduceA\reduceA\reduceA\reduceA

In this section, we introduce our systematization methodology.
We provide intuition into how we select a representative sample of the community's work;
next, we describe the criteria used to systematize knowledge. %

\reduceA\reduceA\reduceA\reduceA\reduceA\reduceA\reduceA
\subsection{Sampling Relevant Work}
\reduceA\reduceA\reduceA\reduceA

\descr{GenomePrivacy.org.} 
We study research on genome privacy from the point of view of privacy-enhancing technologies (PETs) -- specifically, we focus on work using PETs in the context of testing, storing, and/or sharing genomic data.
Therefore, we rely on the website \url{GenomePrivacy.org}, which bills itself as ``the community website for sharing information about research on the technical protection of genome privacy and security.'' 
In the summer of 2017, we retrieved the 197 articles listed on the site,
and grouped them into six canonical categories -- Personal Genomic Testing, Genetic Relatedness Testing, Access and Storage Control, Genomic Data Sharing, Outsourcing, and Statistical Research %
-- from which we selected a sample representing the state of the art for each category.

\descr{Excluding Attack/Quantification Papers.} 
We excluded work on attacks (reviewed in Section~\ref{sec:background}) and privacy quantifications as our main focus is on the use of PETs. 
We refer readers to~\cite{wagner2015genomic} for a comprehensive evaluation of metrics geared to quantify genomic privacy, and to~\cite{wan2017expanding} for game-theoretic approaches to quantify protection in genomic data sharing.
We also did not include recent proposals to address specific attacks in the context of Beacon~\cite{shringarpure2015privacy},
e.g., the work by Raisaro et al.~\cite{raisaro2017addressing} or Wan et al.~\cite{wan2017controlling}, although we note their importance later in Section~\ref{sec:real-life}.

\descr{Selection.} 
To select the list of papers used to drive our systematic analysis (Section~\ref{sec:analysis}), we followed an iterative process.
First, the team selected 45 articles %
considered to represent the state of the art in the six categories,
and later added four more during a revision of the paper in March 2018.
Since it would be impossible to review and systematize all of them in a succinct and seamless manner, we trimmed down the selection to 25 papers (reviewed in Section~\ref{sec:stateoftheart}).
When deciding whether to include one paper over another, the team preferred papers published in venues that are more visible to the privacy-enhancing technologies community or that have been cited significantly more, as they arguably have a stronger influence on the community over time.

Ultimately, the selection covers 
four papers in Personal Genomic Testing, 
three in Genetic Relatedness Testing,
four in Access \& Storage Control, 
six in Genomic Data Sharing, 
six in Outsourcing,
and four in Statistical Research.
Note that two articles appear in both Personal Genomic Testing and Genetic Relatedness Testing categories. 
For completeness, in Section~\ref{sec:stateoftheart}, we also add a citation to the papers from the first list of 49 papers that are not included in the final selection.

\descr{Remarks.} 
We stress that we do not aim to analyze all papers related to genomic privacy in the context of PETs; in fact, our selection is meant to be representative of the state of the art for each category, but not of its breadth or depth.
Rather, we systematize knowledge around genomic privacy protection mechanisms and critically evaluate it.
As a consequence, if we added or replaced one paper with another, the main takeaways would not be considerably altered.

\reduceA\reduceA\reduceA\reduceA\reduceA\reduceA\reduceA
\subsection{Systematization Criteria}\label{sec:criteria}
\reduceA\reduceA\reduceA\reduceA

We now discuss the main axes along which we systematize genome privacy work.
To do so, we elicit a set of criteria designed to address the research questions posed in the introduction.
Inspired by similar approaches in SoK papers~\cite{bonneau2012quest,khattak2016sok}, we choose criteria aiming to capture different aspects related to security, efficiency, and system/data models while being as comprehensive as possible.
These criteria constitute the columns of Table~\ref{tab:table1} (see Section~\ref{sec:analysis}), %
where each row is one of the papers discussed below. 
Specifically, we define the following 9 criteria:

\descrit{1. Data Type.}
We capture the type of genomic data used, e.g., some protocols perform computation on full genomes, or other aspects of the genome such as SNPs or haplotypes. 

\descrit{2. Genomic Assumptions.} 
We elicit whether techniques make any assumptions as to the {\em nature} of the data.
For instance, the processing of sequencing data is not perfect and nucleotides (or even sequences thereof) might be misreported or deleted, while others might be inserted unexpectedly. 
In fact, the error rate percentage across various next-generations sequencers can be as high as 15\%~\cite{goodwin2016coming}. 
As such, the output of modern Illumina sequencing machines (i.e., FASTQ format\footnote{\url{https://help.basespace.illumina.com/articles/descriptive/fastq-files/}}) %
is made of segments of DNA with probabilities associated with the confidence that letters were read correctly. 
This criterion serves to note which of the proposed methodologies take into consideration, or are particularly affected by this.

\descrit{3. Storage Location.} 
We study where genomic data is assumed to be stored. 
We identify three options: 
(i) a personal device, like a mobile phone or a dedicated piece of hardware which is operated by the user, 
(ii) the cloud, from which a user can directly obtain her data or allow a medical facility to obtain it, and 
(iii) institutions (e.g., biobanks and hospitals), which store and are able to process genomic data at will. 
We refer to the latter as {\em Data Controllers}, following GDPR's terminology~\cite{GDPR2016}.

\descrit{4. Use of Third Parties.} 
We determine the presence of third parties, if any, as well as their nature. 
For instance, some protocols may involve key distribution centers and semi-trusted cloud storage providers.

\descrit{5. Long-Term Security.} 
Due to its hereditary nature, the sensitivity of genomic data does not degrade quickly over the years: even access to the genome of a long-deceased individual might still pose a threat to their descendants.
Therefore, we look at the underlying building blocks and the computational assumptions in genome privacy tools and analyze whether or not they can realistically withstand several decades of computational advances.

\descrit{6. Security Assumptions.} 
We study the assumptions made on entities involved, if any. 
For instance, we consider if third parties are assumed not to collude with any other entities.

\descrit{7. Methods.} We report on the main security tools and methods used (e.g., secure multiparty computation, homomorphic encryption).

\descrit{8. Privacy Overhead.} 
We broadly quantify the overhead introduced by the privacy defense mechanisms, compared, whenever possible, to non privacy-preserving versions of the same functionality.
This is a non-trivial task because each sub-area of genome privacy has different goals and each piece of work in that area does not necessarily solve the exact same problem.
Nonetheless, we analyze the complexity of each solution to assess their efficiency in terms of time and storage overhead. 
We report on the lower and upper values of complexity to emphasize how each solution fares against the non-privacy version of the same functionality. 
We do so based on the premise that if the technique imposes orders of magnitude higher overhead than the non-privacy-preserving version, then the overhead is considered to be high, and low otherwise.

\descrit{9. Utility Loss.} 
Finally, we measure the impact of privacy tools on the utility of the system.
Such measurements include the overall flexibility of the proposed work in comparison with the intended task. 
Similar to the privacy overhead criterion, we compare against non-privacy-preserving versions of the same functionality, and quantify utility loss as either low or high.

\descr{Remarks.} 
We do not necessarily report on the specific metrics used in the selected papers (e.g., running times) as (i) not all papers provide metrics, and (ii) similar approaches already appear in prior work (see Section~\ref{sec:sok_vs_survey}). 
Rather, the metrics used in the systematization are designed to support a critical analysis of the PETs invoked to protect genome privacy.

\reduceA\reduceA\reduceA\reduceA\reduceA\reduceA\reduceA\reduceA\reduceA
\section{Representative Papers}\label{sec:stateoftheart}
\reduceA\reduceA\reduceA\reduceA

We now review the papers selected according to the methodology presented in Section~\ref{sec:methodology}.
These papers constitute the rows in Table~\ref{tab:table1}. 
Citations to them appear in red throughout the paper.

\reduceA\reduceA\reduceA\reduceA\reduceA\reduceA\reduceA\reduceA
\subsection{Personal Genomic Testing}
\reduceA\reduceA\reduceA\reduceA

We begin with papers that define privacy-preserving versions of personal genomic tests.
These have a variety of uses, including assessments of a person's predisposition to a disease, determining the best course of treatment, and optimizing drug dosage.
Typically, they involve an individual and a testing facility, and consist of searching for and weighting either short patterns or single nucleotide polymorphisms (SNPs). 
In this context, there are two main privacy-friendly models: (1) one assuming that individuals keep a copy of their genomic data and consent to tests so that only the outcome is disclosed and (2) another involving a semi-trusted party that stores an encrypted copy of the patient's genetic information, and is involved in the interactions.

Baldi et al.~\mycite{{baldi2011countering}} operate in model (1), supporting privacy-preserving searching of mutations in specific genes.
They use authorized private set intersection (APSI)~\cite{de2010practical}, which guarantees that the test is authorized by a regulator (``authorization authority'') and pushes pre-computation offline so that the complexity of the online interaction only depends on the handful of SNPs tested.
It also ensures that the variants which make up the test are kept confidential, as this may pertain to a company's intellectual property.

Ayday et al.~\mycite{{ayday2013protecting}} introduce model (2), letting a Medical Center (MC) perform  private disease susceptibility tests on patients' SNPs, by computing a weighted average of risk factors and SNP expressions. 
In this model patients have their genome sequenced once, through a Certified Institution (CI) that encrypts the SNPs and their positions, and uploads them to a semi-trusted Storage and Processing Unit (SPU). The MC computes the disease susceptibility using cryptographic tools, such as homomorphic encryption and proxy re-encryption.
Also in model (2) is the work by Naveed et al.~\mycite{{naveed2014controlled}}, whereby the CI encrypts genomes using controlled-functional encryption (C-FE), under a public key issued by a central authority, and publishes the ciphertexts.
MCs can then run tests using a one-time function key, obtained by the authority, which corresponds to one specific test and can only be used for that test.

Djatmiko et al.~\mycite{{djatmiko2014secure}} operate in both models (i.e., patients control their data by storing it on a personal device or in the cloud) to support personalized drug dosing (which in this case happens to be Warfarin, a blood thinner). %
The testing facility retrieves data to be evaluated (using private information retrieval~\cite{chor1995private}) and processes it while encrypted. The patient then securely computes the linear combination of test weights (using additively homomorphic encryption), and shows the results to the physician.

\begin{tcolorbox}[title={\small Personal Genomic Testing -- Selected Papers}, boxrule=0.1pt,boxsep=3pt,left=2pt,right=2pt,top=1pt,bottom=1pt]
\small
1. Baldi et al., CCS'11~\mycite{{baldi2011countering}} \\ 
2. Ayday et al., WPES'13~\mycite{{ayday2013protecting}} \\ 
3. Naveed et al., CCS'14~\mycite{{naveed2014controlled}} \\
4. Djatmiko et al., WPES'14~\mycite{{djatmiko2014secure}} 
\end{tcolorbox}

\begin{tcolorbox}[title={\small Additional Papers}, boxrule=0.1pt,boxsep=3pt,left=2pt,right=2pt,top=1pt,bottom=1pt]
\small 
See \cite{troncoso2007privacy}, \cite{blanton2010secure}, \cite{de2013secure}, \cite{shimizu2016efficient}, \cite{mclaren2016privacy} 
\end{tcolorbox}

\reduceA\reduceA\reduceA\reduceA\reduceA\reduceA\reduceA\reduceA\reduceA
\subsection{Genetic Relatedness}\label{subsec:relatedness}
\reduceA\reduceA\reduceA\reduceA
We next look at genetic relatedness, i.e., testing to ascertain genealogy or ancestry of individuals.
Genealogy tests determine whether two individuals are related (e.g., father and child) or to what degree (e.g., they are $n^{th}$ cousins), while, ancestry tests estimate an individual's genetic ``pool'' (i.e., where their ancestors come from).
These tests are often referred to as part of ``recreational genomics'', and are one of the drivers of the DTC market (with 23andMe and AncestryDNA offering them at under \$100).
However, due to the hereditary nature of the human genome, they also raise several privacy concerns~\cite{golden2018}.  
Privacy research in this area aims to support privacy-respective versions of such tests.

Baldi et al.~\mycite{{baldi2011countering}} allow two users, each holding a copy of their genome, to simulate \textit{in vitro} paternity tests based on Restriction Fragment Length Polymorphisms (RFLPs), without disclosing their genomes to each other or third-parties, through the use of private set intersection protocols~\cite{de2012fast}.
He et al.~\mycite{{he2014identifying}} let individuals privately discover their genetic relatives by comparing their genomes to others stored, encrypted, in the same biorepository, using fuzzy encryption~\cite{dodis2004fuzzy} and a novel secure genome sketch primitive, which is used to encrypt genomes using a key derived from the genome itself. %
Finally, Naveed et al.~\mycite{{naveed2014controlled}} rely on C-FE to enable a client to learn certain functions, including paternity and kinship, over encrypted data, using keys obtained from a trusted authority. %

The tools above differ in a few aspects. 
First, \mycite{{baldi2011countering}} assumes individuals obtain and store a copy of their sequenced genome, whereas \mycite{{he2014identifying}} and \mycite{{naveed2014controlled}} operate under the assumption that users will rely on cloud providers.
Second, \mycite{{baldi2011countering}} operates on full genomes, while \mycite{{naveed2014controlled}} supports SNP profiles obtained from DTC genomics companies, with \mycite{{he2014identifying}} requiring individuals' haplotypes.\smallskip

\begin{tcolorbox}[title={\small Genetic Relatdness -- Selected Papers},boxrule=0.1pt,boxsep=3pt,left=2pt,right=2pt,top=1pt,bottom=1pt]
\small
1. Baldi et al., CCS'11~\mycite{{baldi2011countering}} \\ 
2. He et al., Genome Research'14~\mycite{{he2014identifying}} \\
3. Naveed et al., CCS'14~\mycite{{naveed2014controlled}} 
\end{tcolorbox}

\begin{tcolorbox}[title={\small Additional Papers},boxrule=0.1pt,boxsep=3pt,left=2pt,right=2pt,top=1pt,bottom=1pt]
\small \cite{hormozdiari2014privacy}, \cite{de2016relatedness}, \cite{mclaren2016privacy}
\end{tcolorbox}

\reduceA\reduceA\reduceA\reduceA\reduceA\reduceA\reduceA\reduceA\reduceA\reduceA\reduceA
\subsection{Access and Storage Control}
\reduceA\reduceA\reduceA

Next, we discuss results aiming to guarantee secure access to, and storage of, genomic data. 
Karvelas et al.~\mycite{{karvelas2014privacy}} use a special randomized data structure based on Oblivious RAM (ORAM)~\cite{goldreich1996software} to store data while concealing access patterns, using two servers to cooperatively operate the ORAM.
Clients can then query data using a third entity who retrieves encrypted data from the ORAM and instructs the servers to jointly compute functions using secure two-party computation~\cite{yao1986generate}.
Ayday et al.~\mycite{{ayday2014privacy}} present a framework for privately storing, retrieving, and processing SAM files where a CI sequences and encrypts patients' genomes, and also creates the SAM files, storing them encrypted in biorepositories.
Then, MCs using order-preserving encryption~\cite{agrawal2004order} can retrieve data and conduct genetic tests.

Beyond SAM files, genomic data is also stored in BAM (a binary version of SAM) or CRAM files, which allows a lossless compression. Huang et al.~\mycite{huang2016privacy} introduce a Secure CRAM (SECRAM) format, supporting compression, encryption, and selective data retrieval.
SECRAM requires less storage space than BAM, and maintains CRAM's efficient compression and downstream data processing.
Finally, Huang et al.~\mycite{{huang2015genoguard}} focus on long-term security, introducing GenoGuard, a system aiming to protect encrypted genomic data against an adversary who tries to brute-force the decryption key (likely to succeed in 30 years). They rely on Honey Encryption (HE)~\cite{juels2014honey} so that, for any decryption attempt using an incorrect key, a random yet plausible genome sequence is produced.
Overall, we find that security issues in this context are not explored in as much depth as other areas. \smallskip

\begin{tcolorbox}[title={\small Access and Storage Control -- Selected Papers}, boxrule=0.1pt,boxsep=3pt,left=2pt,right=2pt,top=1pt,bottom=1pt]
\small
1. Karvelas et al., WPES'14~\mycite{{karvelas2014privacy}} \\
2. Ayday et al., DPM'14~\mycite{{ayday2014privacy}} \\
3. Huang et al., IEEE S\&P'15~\mycite{{huang2015genoguard}} \\
4. Huang et al., Genome Research'16~\mycite{huang2016privacy} 
\end{tcolorbox}

\begin{tcolorbox}[title={\small Additional Papers},boxrule=0.1pt,boxsep=3pt,left=2pt,right=2pt,top=1pt,bottom=1pt]
\small
\cite{troncoso2007privacy}, \cite{kantarcioglu2008cryptographic}, \cite{blanton2010secure}, \cite{canim2012secure}
\end{tcolorbox}

\reduceA\reduceA\reduceA\reduceA\reduceA\reduceA\reduceA\reduceA
\subsection{Genomic Data Sharing}
\reduceA\reduceA\reduceA

We now discuss results in the context of genomic data sharing, which is an important aspect of hypothesis-driven research.
Consider, for instance, genome wide association studies (GWAS): to elicit robust conclusions on the association between genomic features and diseases and traits, researchers may need millions of samples~\cite{burton2008size}. 
Even if sequencing costs continue to rapidly drop, it is unrealistic to assume that research teams can easily gain access to such a large number of records.
Yet, though there is an interest in data sharing, these sharing initiatives face several obstacles, as 
(1) researchers in isolation may be prevented from (or are hesitant to) releasing data, and 
(2) might only have patients' consent for specific studies at specific institutions.
Therefore, privacy-enhancing methods have been proposed to address these issues.

Kamm et al.~\mycite{{kamm2013new}} present a data collection system where genomic data is distributed among several entities using secret sharing. 
Secure multiparty computation (MPC) is then used to conduct computations on data, privately, supporting secure GWAS across multiple entities, such as hospitals and biobanks.
Xie et al.~\mycite{{xie2014securema}} introduce SecureMA, which allows secure meta-analysis for GWAS. (Meta-analysis is a statistical technique to synthesize information from multiple independent studies~\cite{evangelou2013meta}.)
Their framework generates and distributes encryption/decryption keys to participating entities, encrypts association statistics of each study locally, and securely computes the meta-analysis results over encrypted data. 
Humbert et al.~\mycite{{humbert2014reconciling}} consider the case of individuals willing to donate their genomes to research. %
They quantify the privacy risk for an individual using a global privacy weight of their SNPs and use an obfuscation mechanism that functions by hiding SNPs. \smallskip %

Wang et al.~\mycite{{wang2015efficient}} enable clinicians to privately find similar patients in biorepositories. 
This could be applied, for instance, to find out how these patients respond to certain therapies.
In their paper, similarity is defined as the edit distance~\cite{navarro2001guided}, i.e., the minimum number of edits needed to change one string into another.
Using optimized garbled circuits, they build a genome-wide, privacy-preserving similar patient query system.
This requires participating parties (e.g., medical centers) to agree on a public reference genome and independently compress their local genomes using a reference genome, creating a Variation Call Format (VCF) file. 
The edit distance of two genomes can then be calculated by securely comparing the two VCF files.
Jagadeesh et al.~\mycite{{jagadeesh2017deriving}} enable the identification of causal variants and the discovery of previously unrecognized disease genes while keeping 99.7\% of the participants' sensitive information private using MPC.

Finally, Chen et al.~\mycite{{chen2017princess}} introduce a framework for computing association studies for rare diseases (e.g., the Kawasaki Disease~\cite{khor2011genome}) over encrypted genomic data of different jurisdictions.
They rely on Intel's Software Guard Extensions (SGX), which isolates sensitive data in a protected enclave and allows the secure computation of the results. %

In summary, work in this category focuses on a wide range of problems, from GWAS and meta-analysis to edit distance computation. %
Also, tools primarily build on cryptographic protocols, except for~\mycite{{chen2017princess}}, which relies on SGX.\smallskip

\begin{tcolorbox}[title={\small Genomic Data Sharing -- Selected Papers},boxrule=0.1pt,boxsep=3pt,left=2pt,right=2pt,top=1pt,bottom=1pt]
\small
1. Kamm et al., Bioinformatics'13~\mycite{{kamm2013new}} \\
2. Xie et al., Bioinformatics'14~\mycite{{xie2014securema}} \\
3. Humbert et al., WPES'14~\mycite{{humbert2014reconciling}} \\
4. Wang et al., CCS'15~\mycite{{wang2015efficient}} \\
5. Jagadeesh et al., Science'17~\mycite{{jagadeesh2017deriving}} \\
6. Chen et al., Bioinformatics'17~\mycite{{chen2017princess}} 
\end{tcolorbox}

\begin{tcolorbox}[title={\small Additional Papers},boxrule=0.1pt,boxsep=3pt,left=2pt,right=2pt,top=1pt,bottom=1pt]
\small
\cite{stade2014grabblur}, \cite{zhang2015secure}, \cite{aziz2016secure}, \cite{wang2016healer}, \cite{wan2017expanding}
\end{tcolorbox}

\reduceA\reduceA\reduceA\reduceA\reduceA\reduceA\reduceA\reduceA\reduceA
\subsection{Outsourcing}
\reduceA\reduceA\reduceA\reduceA

At times, research and medical institutions might lack the computational resources required to store or process large genomic datasets locally. 
As such, there is increasing interest in outsourcing data computation to the cloud, e.g., using dedicated services like Google Genomics or Microsoft Genomics.
However, this requires users to trust cloud providers, which raises security and privacy concerns with respect to data of research volunteers and/or patients.
To address these concerns, several solutions have been proposed.
Note that this category relates to the processing of genomic data in a public cloud environment, whereas, the previously discussed Access \& Storage Control category relates to where and how data is stored, {\em regardless} of the location. 

Chen et al.~\mycite{{chen2012large}} propose the use of Hybrid Clouds~\cite{furht2010cloud}, a method that involves both public and private clouds, to enable privacy-preserving read mapping.  
Read mapping is the process of interpreting randomly sampled sequence reads of the human genome. 
Their solution involves two stages: a seeding stage where the public cloud performs exact matching on a large amount of ciphertexts, and an extension stage where the private cloud computes a small amount of computations (such as, edit distance) at the genetic locations found by the seeding stage.
Yasuda et al.~\mycite{{yasuda2013secure}} present a somewhat homomorphic encryption scheme (SWHE) for secure pattern matching using Hamming distance.
More specifically, in this setting physicians supply patients with homomorphic encryption keys who then encrypt their genomic data and upload them to the cloud.
When the physician needs to test whether a certain DNA sequence pattern appears in the patient's genome, the cloud computes the Hamming distance over encrypted DNA sequences and the desired pattern, and sends the (encrypted) result back to the physician. %

Cheon et al.~\mycite{{cheon2015homomorphic}} also use SWHE to calculate the edit distance of two encrypted DNA sequences, allowing data controllers (e.g., patients) to encrypt their genomic data  and upload them to the cloud, which can calculate the edit distance to the reference genome or other encrypted sequences.
Lauter et al.~\mycite{lauter2014private} introduce a leveled homomorphic encryption scheme (LHE) to securely process genomic data in the cloud for various genomic algorithms used in GWAS, such as Pearson and $\chi^2$ Goodness-of-Fit statistical tests.\footnote{LHE is a fully homomorphic encryption scheme variant that does not require bootstrapping but can evaluate circuits with a bounded depth.}
Usually, computation of these statistics require frequencies or counts but, since their scheme cannot perform homomorphic divisions,
~\mycite{lauter2014private} have to modify some of these computations to work with counts only.

Kim and Lauter~\mycite{{kim2015private}} also use SWHE to securely compute minor allele frequencies and $\chi^2$-statistics for GWAS-like applications, over encrypted data, as well as the edit/Hamming distance over encrypted genomic data.
Finally, Sousa et al.~\mycite{{sousa2017efficient}} rely on SWHE and private information retrieval to let researchers search variants of interest in VCF files stored in a public cloud. 
Their solution represents an improvement upon the state of the art in terms of efficiency, however, it suffers from high error rates and poor scalability.

\begin{tcolorbox}[title={\small Outsourcing -- Selected Papers}, boxrule=0.1pt,boxsep=3pt,left=2pt,right=2pt,top=1pt,bottom=1pt]
\small
1. Chen et al., NDSS'12~\mycite{{chen2012large}} \\
2. Yasuda et al., CCSW'13~\mycite{{yasuda2013secure}} \\
3. Lauter et al., LatinCrypt'14 ~\mycite{lauter2014private} \\
4. Cheon et al., FC'15~\mycite{{cheon2015homomorphic}} \\
5. Kim and Lauter, BMC'15~\mycite{{kim2015private}} \\
6. Sousa et al., BMC'17~\mycite{{sousa2017efficient}}
\end{tcolorbox}

\begin{tcolorbox}[title={\small Additional Papers},boxrule=0.1pt,boxsep=3pt,left=2pt,right=2pt,top=1pt,bottom=1pt]
\small
\cite{bos2014private}, \cite{xu2014privacy}, \cite{zhang2015foresee}, \cite{ghasemi2016private} 
\end{tcolorbox}

\reduceA\reduceA\reduceA\reduceA\reduceA\reduceA\reduceA\reduceA\reduceA
\subsection{Statistical Research}\label{sec:statistical}
\reduceA\reduceA\reduceA\reduceA

The last category focuses on attempts to address unintended leakage threats from the disclosure of genomic data statistics, e.g., membership inference attacks discussed in Section~\ref{sec:background}.

A possible defense is through statistical disclosure control, of which differential privacy (DP) is one related approach.
DP enables the definition of private functions that are free from inferences, providing as accurate query results as possible, while minimizing the chances for an adversary to identify the contents of a statistical database~\cite{dwork2006calibrating}.

Johnson and Shmatikov~\mycite{johnson2013privacy} point out that it is inherently challenging to use DP techniques for GWAS, since these methods output correlations between SNPs while the number of outputs is far greater than that of the inputs (i.e., the number of participants).
In theory, it is possible to limit the number of available outputs and provide results with adequate accuracy~\cite{bhaskar2010discovering,fienberg2011privacy}.
In practice, however, this requires researchers to know beforehand what to ask (e.g., the top-$k$ most significant SNPs),
which is usually infeasible because finding all statistically significant SNPs is often the goal of the study.
To address this issue,~\mycite{johnson2013privacy} define a function based on the exponential mechanism, which adds noise and works for arbitrary outputs.
Their mechanism allows researchers to perform exploratory analysis, including computing in a differentially private way: 
i) the number and location of the most significant SNPs to a disease, 
ii) the $p$-values of a statistical test between a SNP and a disease, 
iii) any correlation between two SNPs, and 
iv) the block structure of correlated SNPs.

Uhlerop et al.~\mycite{uhlerop2013privacy}  aim to address Homer's attack~\cite{homer2008resolving} using a differentially private release of aggregate GWAS data, supporting a computation of differentially private $\chi^2$-statistics and $p$-values, and provide a DP algorithm for releasing these statistics for the most relevant SNPs. They also support the release of averaged minor allele frequencies (MAFs) for the cases and for the controls in GWAS.
Tram{\`e}r et al.~\mycite{tramer2015differential} build on the notion of Positive Membership Privacy~\cite{li2013membership} and introduce a weaker adversarial model, %
also known as Relaxed DP, in order to achieve better utility by identifying the most appropriate adversarial setting and bounding the adversary's knowledge.
Finally, Backes et al.~\mycite{{backes2016privacy}} study privacy risks in epigenetics, specifically, showing that blood-based microRNA expression profiles can be used to identify individuals in a study, and propose a DP mechanism to enable privacy-preserving data sharing.

\begin{tcolorbox}[title={\small Statistical Research -- Selected Papers},boxrule=0.1pt,boxsep=3pt,left=2pt,right=2pt,top=1pt,bottom=1pt]
\small
1. Johnson and Shmatikov, KDD'13~\mycite{johnson2013privacy} \\
2. Uhlerop et al., JPC'13~\mycite{uhlerop2013privacy} \\
3. Tram{\`e}r et al., CCS'15~\mycite{tramer2015differential} \\
4. Backes et al., USENIX''16~\mycite{{backes2016privacy}} \\[0.5ex]
Papers on statistical methods are not reported in Table~\ref{tab:table1} because the systematization criteria do not apply, but, for context, they are discussed in Section~\ref{sec:cost_of_privacy}. 
\end{tcolorbox}

\begin{tcolorbox}[title={\small Additional Papersk},boxrule=0.1pt,boxsep=3pt,left=2pt,right=2pt,top=1pt,bottom=1pt]
\small
\cite{yu2014scalable}, \cite{zhao2014choosing}, \cite{jiang2014community}, \cite{wang2016healer}, \cite{simmons2016realizing}.
\end{tcolorbox}
\reduceA\reduceA\reduceA\reduceA

\begin{table*}[p]
\small
\captionsetup{singlelinecheck = false, justification = raggedright} %
\centering
\newcolumntype{R}[2]{%
  >{\adjustbox{angle=#1,lap=\width-(#2)}\bgroup}%
  l%
  <{\egroup}%
}
\newcommand*\rot{\multicolumn{1}{R{35}{1em}}} %
\begin{tabu} to \textwidth {X[1.6]XXX[1.5]XXXX[2.1]X[1.5]X}
\toprule
                                    & \rot{\bf Data Type}   & \rot{\bf Genomic Assumptions}   & \rot{\bf Storage Location}  & \rot{\bf Long-Term Security}  & \rot{\bf Third Parties}   & \rot{\bf Security Assumptions}    & \rot{\bf Methods}       & \rot{\bf Privacy Overhead}    & \rot{\bf Utility Loss}    \\ \midrule
\multicolumn{10}{l}{\bf Personal Genomic Testing}                                                                                                                                                                                                                                                               \\ \midrule
~\mycite{{baldi2011countering}}     & FSG                   & Yes                             & User                        & No                            & AA                        & SH, NC                            & A-PSI                 & HSO, LTO                      & Low                       \\
~\mycite{{ayday2013protecting}}     & SNP                   & No                              & Cloud                       & No                            & SPU                       & SH, NC                            & Paillier, Proxy       & HSO, LTO                      & Low                       \\
~\mycite{{djatmiko2014secure}}      & SNP                   & No                              & User/Cloud                  & No                            & N/CS                      & SH                                & Paillier, PIR         & LSO, LTO                      & high                      \\
~\mycite{{naveed2014controlled}}    & SNP                   & No                              & Cloud                       & No                            & CA, CS                    & SH, NC                            & C-FE                  & LSO, LTO                      & Low                       \\ \midrule
\multicolumn{10}{l}{\bf Genetic Relatedness Testing}                                                                                                                                                                                                                                                            \\ \midrule
~\mycite{{baldi2011countering}}     & FSG                   & Yes                             & User                        & No                            & No                        & SH                                & PSI-CA                & LSO, LTO                      & High                      \\
~\mycite{{he2014identifying}}       & SNP                   & No                              & Cloud                       & No                            & CS                        & SH                                & Fuzzy                 & LSO, LTO                      & High                      \\
~\mycite{{naveed2014controlled}}    & SNP                   & No                              & Cloud                       & No                            & CA, CS                    & SH, NC                            & C-FE                  & LSO, LTO                      & Low                       \\ \midrule
\multicolumn{10}{l}{\bf Access \& Storage Control}                                                                                                                                                                                                                                                              \\ \midrule
~\mycite{{karvelas2014privacy}}     & FSG                   & No                              & Cloud                       & No                            & CS                        & SH, NC                            & ElGamal, ORAM         & HSO, HTO                      & Low                       \\
~\mycite{{ayday2014privacy}}        & FSG                   & No                              & Cloud                       & No                            & CS, MK                    & SH, NC                            & OPE                   & HSO, LTO                      & Low                       \\
~\mycite{{huang2015genoguard}}      & FSG                   & Yes                             & Cloud                       & Yes                           & CS                        & SH                                & HoneyEncr             & LSO, HTO                      & High                      \\
~\mycite{{huang2016privacy}}        & FSG                   & No                              & User/Cloud                  & No                            & No                        & SH                                & OPE                   & LSO, LTO                      & Low                       \\ \midrule
\multicolumn{10}{l}{\bf Genomic Data Sharing}                                                                                                                                                                                                                                                                   \\ \midrule
~\mycite{{kamm2013new}}             & SNP                   & No                              & Cloud                       & Yes                           & CS                        & SH, NC                            & SecretSharing         & LSO, HTO                      & High                      \\
~\mycite{{xie2014securema}}         & SNP                   & No                              & DataController              & No                            & KDC                       & SH, NC                            & Paillier, MPC         & LSO, LTO                      & High                      \\
~\mycite{{humbert2014reconciling}}  & SNP                   & No                              & Cloud                       & No                            & No                        & SH                                & DataSuppr             & ---, LTO                      & Low                       \\ 
~\mycite{{wang2015efficient}}       & VCF                   & No                              & DataController              & No                            & No                        & SH                                & MPC                   & LSO, LTO                      & High                      \\
~\mycite{{chen2017princess}}        & SNP                   & No                              & DataController              & No                            & No                        & SGX                               & SGX                   & LSO, LTO                      & High                      \\ 
~\mycite{{jagadeesh2017deriving}}   & FSG                   & No                              & User                        & No                            & No                        & SH                                & MPC                   & Varies$^*$                    & Low                       \\ \midrule                                                                                                  
\multicolumn{10}{l}{\bf Outsourcing}                                                                                                                                                                                                                                                                            \\ \midrule
~\mycite{{chen2012large}}           & FSG                   & No                              & Cloud                       & No                            & CS                        & SH                                & Hash             		& HSO, HTO                      & Low                       \\                     
~\mycite{{yasuda2013secure}}        & FSG                   & No                              & Cloud                       & No                            & CS                        & SH                                & SWHE                  & LSO, LTO                      & High                      \\
~\mycite{lauter2014private}         & SNP                   & No                              & Cloud                       & No                            & CS                        & SH                                & LHE                   & LSO, LTO                      & High                      \\
~\mycite{{cheon2015homomorphic}}    & FSG                   & No                              & Cloud                       & No                            & CS                        & SH                                & SWHE                  & LSO, HTO                      & High                      \\
~\mycite{{kim2015private}}          & FSG                   & No                              & Cloud                       & No                            & CS                        & SH                                & SWHE                  & LSO, HTO                      & Low                       \\ 
~\mycite{{sousa2017efficient}}      & VCF                   & No                              & Cloud                       & No                            & CS                        & SH                                & SWHE, PIR             & ---, LTO                      & High                      \\ %
\bottomrule
\end{tabu}
\begin{tablenotes}
\small
  \item \hspace*{-0.08cm}\textbf{\emph{Third Parties}}: CS: Cloud Storage, SPU: Storage \& Processing Unit, AA: Authorization Authority, CA: Central Authority, KDC: Key Distribution Center, MK: Masking \& Key Manager, No: No Third Party \vspace{0.05cm}
  \item \hspace*{-0.08cm}\textbf{\emph{Data}}: FSG: Fully Sequenced Genome, SNP: SNPs, Hap: Haplotypes, VCF: Variation Call Format \vspace{0.05cm}
  \item \hspace*{-0.08cm}\textbf{\emph{Methods}}: SWHE: Somewhat Homomorphic Encryption, LHE: Leveled Homomorphic Encryption, Fuzzy: Fuzzy Encryption, PSI-CA: Private Set Intersection Cardinality, A-PSI: Authorized Private Set Intersection, C-FE: Controlled Functional Encryption, HoneyEncr: Honey Encryption, OPE: Order-Preserving Encryption, MPC: Secure Multiparty Computation, PIR: Private Information Retrieval, SGX: Software Guard Extensions \vspace{0.05cm}
  \item \hspace*{-0.08cm}\textbf{\emph{Security Assumptions}}: NC: No Collusions, SGX: Software Guard Extensions \vspace{0.05cm}
  \item \hspace*{-0.08cm}\textbf{\emph{Privacy Overhead}}: LSO: Low Storage Overhead, HSO: High Storage Overhead, LTO: Low Time Overhead, HTO: High Time Overhead \vspace{0.05cm}
  \item \hspace*{-0.08cm}\textbf{$^*$\emph{Varies}}: Depends on the Input Size
  \vspace*{-0.1cm}
\end{tablenotes}
\noindent\makebox[\linewidth]{\rule{\textwidth}{1pt}}
\vspace{-0.5cm}
\caption{A systematic comparison of the representative genomic privacy methodologies. The rows represent each work and the columns represent the list of criteria we apply for assessment purposes.}
\label{tab:table1}
\vspace{-0.2cm}
\end{table*}

\reduceA\reduceA\reduceA\reduceA\reduceA\reduceA\reduceA
\section{Systematic Analysis}\label{sec:analysis}
\reduceA\reduceA\reduceA\reduceA

This section reports on the systematic analysis of privacy-enhancing technologies in the genomics context, as they stand today, building on the methodology and the research results discussed in Section~\ref{sec:methodology} and~\ref{sec:stateoftheart}, respectively.
We drive our discussion from Table~\ref{tab:table1}, which, in addition to providing an overview of the community's work, concisely summarizes the results of the analysis. 
It further enables a discussion on insights, research gaps, as well as challenges to certain assumptions. 
In the process, we highlight a list of ten technical challenges.

\reduceA\reduceA\reduceA\reduceA\reduceA\reduceA\reduceA
\subsection{The Issue of Long-Term Security}\label{sec:long}
\reduceA\reduceA\reduceA\reduceA

The longevity of security and privacy threats stemming from the disclosure of genomic data is substantial for several notable reasons. First, access to an individual's genome allows an adversary to deduce a range of genomic features that may also be relevant for her descendants, possibly several generations down the line. Thus, the sensitivity of the data does not necessarily degrade quickly, even after its owner has deceased.
Moreover, the full extent of the inferences one can make from genomic data is still not clear, as researchers are still studying and discovering the relationship between genetic mutations and various phenomena.

These issues also imply that the threat model under which a volunteer decides to donate their genome to science, or have it tested by a DTC company, is likely to change in the future. 
As a consequence, the need or desire to conceal one's genetic data might evolve. 
For instance, a family member may decide to enter politics, or a country's political landscape shifts toward supporting racist ideologies aimed to discriminate against members of a certain ancestral heritage. %

\descr{Inadequacy of Standard Cryptographic Tools.} 
We find that the vast majority of genome privacy solutions rely on cryptographic tools, yet, they are not fit for purpose if long-term security is to be protected.
Modern cryptosystems assume that the adversary is computationally bounded, vis-\`a-vis a ``security parameter.''
Suggestions for appropriate choices of the value for this parameter, and resulting key sizes, are regularly updated by the cryptography community, however, assuming at most the need for security for thirty to fifty years~\cite{enisa}.
While this timeframe is more than adequate in most cases (e.g., classified documents get regularly de-classified and financial transactions/records become irrelevant), it may not be in the case of genomic data.

In theory, one could increase key lengths indefinitely, but, in practice, this is not possible for all cryptosystems,
e.g.,  the block and stream ciphers available today are only designed to work with keys up to a certain length, and libraries implementing public-key cryptography also impose a limit on key sizes.
Furthermore, flaws in cryptosystems considered secure today may be discovered (as happened, recently, with RC4 or SHA-1), and quantum computing might eventually become a reality~\cite{ibm2018}.

\descr{Implications.} 
Naturally, the issue of long-term security affects different genome privacy solutions in different ways.
For instance, if genomic information is stored in an encrypted form and processed by a specialized third entity, such as the SPU in~\mycite{{ayday2013protecting}}, then a malicious or compromised entity likely has multiple chances over time to siphon encrypted data off and succeed in decrypting it in the future.
This is also the case in settings where biobanks store patients' encrypted SAM files~\mycite{{ayday2014privacy}}
or in the context of secure outsourcing solutions, where genomic information is offloaded and encrypted, to a cloud provider.
On the other hand, if encrypted data is only exchanged when running cryptographic protocols, but not stored long-term elsewhere (as in~\mycite{baldi2011countering,djatmiko2014secure,xie2014securema}), then the adversary has a more difficult task.
Nonetheless, long-term security compromise is still possible, even by an eavesdropping adversary and even if the protocol run is super-encrypted using TLS. %
In fact, documents leaked by Edward Snowden revealed that the NSA has tapped submarine Internet cables and kept copies of encrypted traffic~\cite{submarine2013,verge2014}. %

\descr{Possible Countermeasures.} 
Ultimately, genome privacy literature has not sufficiently dealt with long term security.
In fact, only the work by Huang et al.~\mycite{{huang2015genoguard}} attempts to do so, relying on Honey Encryption to encrypt and store genomic data.
Though a step in the right direction, this technique only serves as a storage mechanism and does not support selective retrieval of genomic information, testing over encrypted data, and data sharing.
Moreover, it suffers from several security limitations. 
Specifically, while their solution provides information-theoretic guarantees (and long-term security), their threat model needs to account for possible side-channel attacks.
This is because, if the adversary knows some of the target's physical traits (e.g.,  hair color or gender),  then it can easily infer that the decryption key she is using is not the correct one. The authors attempt to address this issue by making their protocol phenotype-compatible for the cases of gender and  ancestry,  but  there  are  many  other traits in the human genome that possess probabilistic genotype-phenotype associations~\cite{lippert2017identification} thus making it very hard to fully address.

Cryptosystems providing information theoretic security could help, as they are secure even when the adversary has unlimited computing power. 
Unfortunately, they require very large keys and do not support the homomorphic properties needed to perform typical requirements for genomic data (e.g., testing or sharing).
Work relying on secret sharing (e.g., \mycite{{kamm2013new}}) is somewhat an exception, in that it can provide information-theoretic guarantees. 
However, for secret sharing to work, one needs non-colluding entities, which is a requirement that is not always easy to attain (see Section~\ref{sec:sec-lim}).

\begin{tcolorbox}[title={P1. Long-Term Security}, boxrule=0.1pt,boxsep=3pt,left=2pt,right=2pt,top=1pt,bottom=1pt]
An individual's genomic sequence does not change much over time, thus, the sensitivity of the information it conveys may not diminish. However, cryptographic schemes used by PETs in genomics guarantee security only for 20-30 years.
\end{tcolorbox}

\subsection{Security Limitations}\label{sec:sec-lim}

Next, we focus on a number of security assumptions made by some genome privacy protocols.

\descr{Semi-honest Adversaries.} 
All papers listed in Table~\ref{tab:table1}, as well as the vast majority of genome privacy solutions, consider only semi-honest security. Rare exceptions are represented by possible {\em extensions} to~\cite{barman2015privacy,baldi2011countering}.
This is because solutions in this model are significantly easier to instantiate and yield computation and communication complexities that are orders of magnitude lower than in the malicious model. 

However, security in the semi-honest model assumes that the parties do not deviate from the protocol and fails to guarantee correctness (i.e., a corrupted party cannot cause the output to be incorrectly distributed) or input independence (i.e., an adversary cannot make its input depend on the other party's input)~\cite{hazay2010efficient}.
Moreover, in the semi-honest model, parties are assumed to not alter their input.
In practice, these requirements impose important limitations on the real-world security offered by genome privacy solutions. 
Specifically, it might not suffice to ensure that protocols only disclose the outcome of a test to a testing facility or provide hospitals with only information about common/similar patients. 
Indeed, this makes no guarantees as to whether the contents of the test or the patient information has not been maliciously altered or inflated. 
Additionally, the privacy layer makes it more difficult and, at times, impossible, to verify the veracity of the inputs. 

\begin{tcolorbox}[title={P2. Malicious Security}, boxrule=0.1pt,boxsep=3pt,left=2pt,right=2pt,top=1pt,bottom=1pt]
Most genome privacy solutions are designed for settings where the adversaries are considered to be honest-but-curious as opposed to malicious, which may impose limitations on real-world security.
\end{tcolorbox}

\reduceA\reduceA
\descr{Non-Collusion.} 
We also observe that a number of solutions that involve third parties (e.g., for storage and processing encrypted genomic data~\mycite{{ayday2013protecting}}, issuing keys~\mycite{{naveed2014controlled}}, and authorizing tests~\mycite{baldi2011countering,naveed2014controlled}) assume that these parties do not collude with other entities.
Such an assumption has implications of various degrees in different contexts.
For instance,~\mycite{{naveed2014controlled}} assumes that a central authority (CA) is trusted to issue policies (i.e., generating one-time decryption keys, allowing researchers to access a specific genome for a specific case). 
The CA is expected to be operated by some established entity such as the FDA, so that one can likely assume it has no incentive to misbehave (unless compromised).
Similarly, protocols supporting large-cohort research, like the one in~\mycite{{xie2014securema}}, involve medical centers with little or no economic incentive to collude, and violate patients' privacy.

On the other hand, in some cases, non-collusion might be harder to enforce, while the consequences of collusion might be serious.
For instance, the framework in~\mycite{{ayday2013protecting}} supports private disease susceptibility tests, and involves three entities: (i) the Patient, (ii) the MC, which administers the tests, and (iii) the SPU, which stores patients' encrypted SNPs.
Data stored at the SPU is anonymized. 
However, if the SPU and MC collude, then the SPU can re-identify patients.
Moreover, the MC's test specifics must be considered sensitive (e.g., a pharmaceutical company's intellectual property), otherwise there would be no point in performing {\em private} testing. 
This is because one could simply tell the patient/SPU which SNPs to analyze and run the test locally. 
However, patient and SPU collusion implies that confidentiality of the MC's test would be lost.
Also, solutions that assume third-party cloud storage providers do not collude with testing facilities, such as~\mycite{{karvelas2014privacy}}, are limited to settings where one can truly exclude financial or law enforcement disincentives to collusion. 

\begin{tcolorbox}[title={P3. Non-Collusion Assumption}, boxrule=0.1pt,boxsep=3pt,left=2pt,right=2pt,top=1pt,bottom=1pt]
Some genome privacy solutions involve a collection of entities. These solutions further assume that the entities do not collude with each other, which may be difficult to enforce or verify.
\end{tcolorbox}

\noindent\textbf{Trusted Hardware.} 
Other assumptions relate to secure hardware, like SGX, which isolates sensitive data into a protected enclave, thus supporting secure computation of the results, even if the machine is compromised.
For instance,~\mycite{{chen2017princess}} relies on secure hardware to enable institutions to securely conduct computations over encrypted genomic data. 
However, side-channel attacks have been recently demonstrated to be possible~\cite{brasser2017software,hahnel2017high} and the full extent of SGX security has yet to be explored.

\begin{tcolorbox}[title={P4. Trusted Hardware}, boxrule=0.1pt,boxsep=3pt,left=2pt,right=2pt,top=1pt,bottom=1pt]
Some genome privacy solutions rely on trusted hardware, such as SGX. However, the security of such hardware is not yet fully understood and side-channel attacks may limit the security of these solutions.
\end{tcolorbox}

\reduceA\reduceA\reduceA\reduceA\reduceA\reduceA\reduceA\reduceA
\subsection{The Cost of Protecting Privacy}\label{sec:cost_of_privacy}
\reduceA\reduceA\reduceA

Genome privacy research mostly focuses on providing privacy-preserving versions of genomics-related functionalities (e.g., testing, data processing, and statistical research).
While some of these functionalities are already in use (e.g., personal genomic tests offered by DTC companies, data sharing initiatives), others do not yet exist, at least in the way the genome privacy community has envisioned them. 
For instance, some investigations assume that individuals will soon be able to obtain a copy of their fully sequenced genome~\mycite{{baldi2011countering}} or
that we will be able to create an infrastructure and a market with dedicated providers to store and process genomic data for third-party applications~\mycite{ayday2013protecting,karvelas2014privacy}.
Table~\ref{tab:table1} attempts to evaluate the overhead incurred by privacy protection on efficiency and scalability, by comparing to that of supporting its functionality in a non privacy-preserving way.
Similarly, we measure the loss in utility and flexibility.

\descr{Privacy Overhead.} 
We observe that high privacy overhead is linked to the use of expensive cryptographic tools, (e.g., ORAM, Paillier, and SWHE).
On the one hand, we can assume that some might become increasingly efficient in the future, thanks to breakthroughs in circuit optimization~\cite{songhori2015tinygarble}. %
Moreover, the efficiency of fully homomorphic encryption has improved several orders of magnitude over the last couple of years~\cite{register}. %

On the other hand, the characteristics of the privacy properties under consideration intrinsically make the problem harder. 
As a result, it is less likely that efficiency will eventually improve in the foreseeable future. 
For instance, in personal genomic testing, a basic privacy property is concealing which parts of the genome are being tested. 
This implies that every single part needs to be touched, even if the test only needs to examine a few positions.
Some solutions~\mycite{ayday2013protecting,baldi2011countering} partially address this issue through means of pre-computation. 
This is accomplished by encrypting genomic data so that it can be privately searched. 
However, the ciphertext still needs to be transferred in its entirety.
Another example is in the context of genealogy testing, where the goal is to find relatives and distant cousins~\mycite{{he2014identifying}}. 
Accomplishing this in the encrypted domain requires the presence of a central, non-colluding authority, which, as discussed above, is not always feasible. 
A similar situation arises in the context of data sharing: while secure two-party computation can efficiently support pairwise privacy-friendly information sharing, these do not scale well to a large number of participating entities.

\begin{tcolorbox}[title={P5. Privacy Overhead}, boxrule=0.1pt,boxsep=3pt,left=2pt,right=2pt,top=1pt,bottom=1pt]
Some technical genome privacy solutions rely on cryptographic tools (e.g., homomorphic encryption, garbled circuits, or ORAM). These often come with non-negligible computation and communication overheads.
\end{tcolorbox}

\reduceA
\descr{Data Representation.} 
In Table~\ref{tab:table1}, we capture the type of data each solution works with. 
For instance, some protocols operate on SNPs (e.g., \mycite{ayday2013protecting,naveed2014controlled}),  others support FSGs (e.g., \mycite{baldi2011countering,karvelas2014privacy}).
On the one hand, working with FSGs means that researchers and clinicians can consider the genome as a whole, supporting various services, such as research and testing relevant to rare genetic disorders. 
On the other hand, this might be challenging, especially in the ciphertext domain. 
For instance, genome sequencing is still not an error-free process: nucleotides are often misread by the sequencing machines, especially when operating at lower costs. %
Additionally, deletions/insertions of nucleotides are not uncommon and the exact length of the human genome may vary among individuals. %
Handling with such issues is easier in-the-clear than in the ciphertext domain.

In some cases, solutions like~\mycite{{baldi2011countering}} assume simplified formats where the genome is stored and processed as a long vector of nucleotides along with their exact position.
Yet, when errors, deletions, or insertions are not identified before encryption, the accuracy of testing will dramatically reduce (testing in~\mycite{{baldi2011countering}} requires access to specific positions of a vector containing all nucleotides in the genome, thus, if an unidentified insertion or a deletion occurs, the position would shift and the test would not work).
Also, important metadata contained in standard formats (such as, SAM, BAM, and FASTQ) is lost in such a custom representation.
(Note that all of the selected papers use \textit{only} the data type reported in Table~\ref{tab:table1} as an input; 
however, if a tool works with fully sequenced genomes (FSG), it can also support other formats (e.g., one can extract SNPs from an FSG). %
Finally, a non-negligible portion of genome privacy investigations requires systems to change the way they store and process genomic data, which can create challenging hurdles to adoption.

\begin{tcolorbox}[title={P6. Data Representation}, boxrule=0.1pt,boxsep=3pt,left=2pt,right=2pt,top=1pt,bottom=1pt]
Some genome privacy solutions make data representation assumptions, e.g., introducing a custom or simplified data format, not taking into account sequencing errors, removing potentially useful metadata. 
\end{tcolorbox}

\reduceA
\descr{Utility Loss.} Finally, Table~\ref{tab:table1} suggests that, in many instances, the loss in utility, when compared to the corresponding functionality in a non-privacy-preserving setting, is high overall. 
For instance, this may arise due to data representation assumptions discussed above, or because the functionality needs to be adapted for privacy-enhancing tools to work.
Consider that the edit distance algorithm in~\mycite{{cheon2015homomorphic}} can only support small parameters (thus, short sequences), while in~\mycite{lauter2014private} algorithms like Estimation Maximization need to be modified.

Overall, privacy protection inevitably yields a potential loss in utility, as the exact amount of information that should be disclosed needs to be determined ahead of time and rigorously enforced.
As such, a clinician performing a test on the patient's genome loses the freedom to look at whichever data she might deem useful for diagnostic purposes. 
Similarly, a researcher looking for relevant data in large-cohorts might be limited as to what can be searched.
A related consideration can be made with respect to data portability across different institutions.
For instance, if a patient's genomic information is encrypted and stored in a hospital's specialized unit~\mycite{{ayday2013protecting}}, and the patient is traveling or visits another medical facility, it may be significantly harder to access and use her data.

\begin{tcolorbox}[title={P7. Utility Loss (Functionality)}, boxrule=0.1pt,boxsep=3pt,left=2pt,right=2pt,top=1pt,bottom=1pt]
Some genome privacy solutions may result in a loss of utility in terms of the functionality for clinicians and researchers. For instance, privacy tools might limit flexibility, portability, and/or access to data. 
\end{tcolorbox}

\reduceA
\descr{The Challenges of Statistical Research.}\label{sec:dp}
In Section~\ref{sec:stateoftheart}, we reviewed certain efforts~\mycite{johnson2013privacy,uhlerop2013privacy,tramer2015differential,backes2016privacy} to
achieve genome privacy using differentially private mechanisms.
The use of DP in the context of statistical research is limited by the inherent trade-off between privacy and utility.
DP mechanisms aim to support the release of aggregate statistics while minimizing the risks of re-identification attacks.
In this context, every single query yields some information leakage regarding the dataset, and, as the number of queries increases, so does the overall leakage.
Therefore, to maintain the desired level of privacy, one has to add more noise with each query, which can degrade the utility of the mechanism.
The privacy-vs-utility trade-off is a common theme in DP, %
although in many settings genomic data can present unique characteristics with respect to its owner, thus further compounding the problem.
This challenge is exemplified in a case study by Fredrikson et al.~\cite{fredrikson2014privacy}, which focused on a DP release of models for personalized Warfarin dosage. 
In this setting, DP is invoked to guarantee that the model does not leak which patients' genetic markers were relied upon to build the model. 
They show that, to effectively preserve privacy, the resulting utility of the model would be so low that patients would be at risk of strokes, bleeding events, and even death. 

However, in some settings, privacy and utility requirements might not be fundamentally at odds, and could be balanced with an appropriate privacy budget. 
For instance, \mycite{uhlerop2013privacy} show that adding noise directly to the $\chi^2$-statistics, rather than on the raw values, yields better accuracy, while~\mycite{johnson2013privacy} demonstrate that the accuracy of the private statistics increases with the number of patients in the study.
Also, Tram{\`e}r et al.'s techniques~\mycite{tramer2015differential} can achieve higher utility than~\mycite{uhlerop2013privacy,johnson2013privacy} by bounding the adversary's background knowledge. 
Moreover, it has also shown to be challenging to convince biomedical researchers, who are striving to get the best possible results, to accept a non-negligible toll on utility~\cite{mclaren2016privacy}.

\begin{tcolorbox}[title={P8. Utility Loss (Statistical Research)}, boxrule=0.1pt,boxsep=3pt,left=2pt,right=2pt,top=1pt,bottom=1pt]
Some genome privacy solutions rely on differential privacy, i.e., introducing noise to the data which yields a potential non-negligible loss in utility.
\end{tcolorbox}
\reduceA\reduceA

\reduceA\reduceA\reduceA\reduceA\reduceA\reduceA\reduceA
\subsection{Real-Life Deployment}\label{sec:real-life}
\reduceA\reduceA\reduceA\reduceA\reduceA

\descr{Relevance to Current Genomics Initiatives.} 
An important aspect of the genome privacy work to date is its relevance to current genomics initiatives and whether solutions introduced can be used in practice, to enhance the privacy of their participants.
At the moment, these initiatives deal with privacy by relying on access control mechanisms and informed consent, but ultimately require participants to voluntarily agree to make their genomic information available to any researchers who wish to study it. 
Surprisingly, we only came across one solution that could be leveraged for this purpose, although it would require infrastructure changes. 
Specifically, the controlled-functional encryption (C-FE) protocol presented in~\mycite{{naveed2014controlled}} would allow participants' data to be encrypted under a public key issued by a central authority. 
This would allow researchers to run tests using a one-time function key, obtained by the authority, which corresponds to a specific test and can only be used for that purpose.
This means that the authority would need to issue a different key for each individual, for every request, and for every function.
Unfortunately, this is not practical in the context of large-scale research involving millions of participants and hundreds (if not thousands) of researchers.

However, there actually is work aiming to address some attacks against {\em data sharing} initiatives, e.g., membership inference against the Beacon network~\cite{shringarpure2015privacy} (see Section~\ref{sec:sok_vs_survey}).
To address this attack, Raisaro et al.~\cite{raisaro2017addressing} propose that a Beacon should answer positively only if the individuals containing the queried mutation are more than one. 
They also propose hiding a portion of unique alleles using a binomial distribution and providing false answers to queries targeting them,
or imposing a query budget.
Wan et al.~\cite{wan2017controlling} measure the discriminative power of each single nucleotide variant (SNV) in identifying a target in a Beacon, and flip the top SNVs according to this power, 
measuring the effects on privacy and utility. 
However, both solutions make important trade-offs with respect to utility. %
More specifically,~\cite{raisaro2017addressing} alters or limits the responses of the network, while~\cite{wan2017controlling}, acknowledging that utility loss is unavoidable, provides ways to calculate (but not solve) this trade-off.

\begin{tcolorbox}[title={P9. Relevance to Genomics Initiatives}, boxrule=0.1pt,boxsep=3pt,left=2pt,right=2pt,top=1pt,bottom=1pt]
Current genomics initiatives -- e.g., All of Us or Genomics England -- primarily address privacy by relying on access control mechanisms and informed consent. In the meantime, we lack technical genome privacy solutions that can be applied to such initiatives. 
\end{tcolorbox}

\reduceA\reduceA\reduceA
\descr{Relevance to Private Market.} 
We also reason about the applicability of PETs to the genomics private market. 
As a case study, we consider DTC products for health reports and genetic relatedness testing, where the potential drawbacks mentioned earlier (e.g., in terms of utility loss or computational overhead) might be less relevant than in clinical settings.

Today, companies like AncestryDNA and 23andMe provide cheap and seamless services, while investing substantial resources on improving user experience. 
Moreover, as their customer base grows, they can discover and match a greater number of relatives, as well as increase the accuracy of their models using metadata provided by the users.
In return, these companies can profit from monetizing customers' data, e.g., by helping researchers recruit participants for research studies or providing pharmaceutical companies with access to data at a certain price~\cite{wired2015}. 
However, without access to data, their business model is not viable. 
This is because deploying privacy-preserving testing would require the presence of entities that are willing to operate these services with minimal data monetization prospects. 
Notably, this is not a new issue in privacy, and similar considerations can be raised about research on privacy-preserving social networking or cloud computing, which has also struggled with adoption.

\begin{tcolorbox}[title={P10. Relevance to Private Market}, boxrule=0.1pt,boxsep=3pt,left=2pt,right=2pt,top=1pt,bottom=1pt]
Direct-to-consumer genetic testing companies monetize customers' data, and/or use it for research studies. As such, genome privacy solutions for personal genome tests may lack a viable business model.
\end{tcolorbox}

\section{Experts' Opinions}\label{sec:directions}
\begin{figure*}[t]
\centering
\subfigure[Importance]{\includegraphics[width=0.50\textwidth]{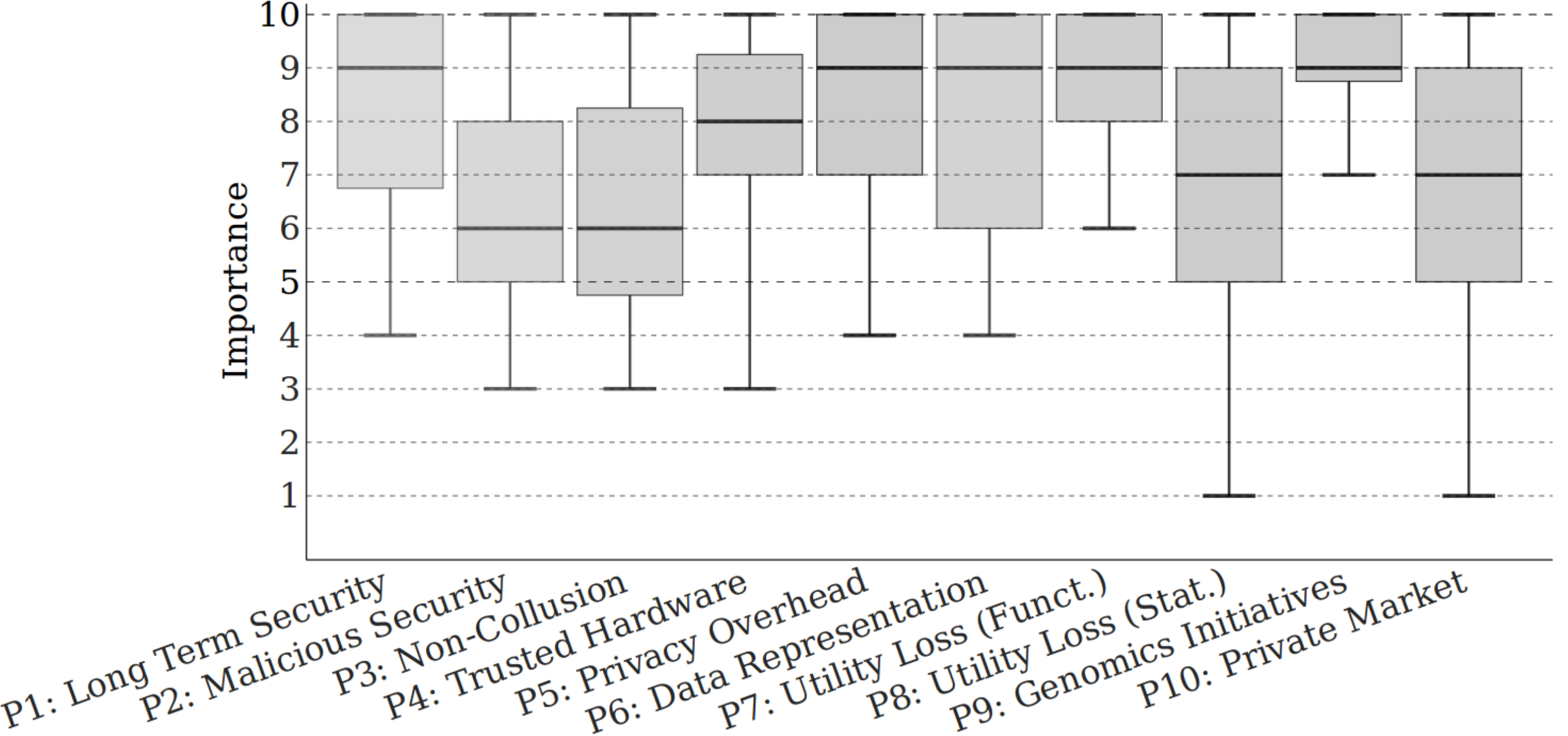}\label{fig:importance}}
\hspace*{-0.4cm}
\subfigure[Difficulty]{\includegraphics[width=0.5\textwidth]{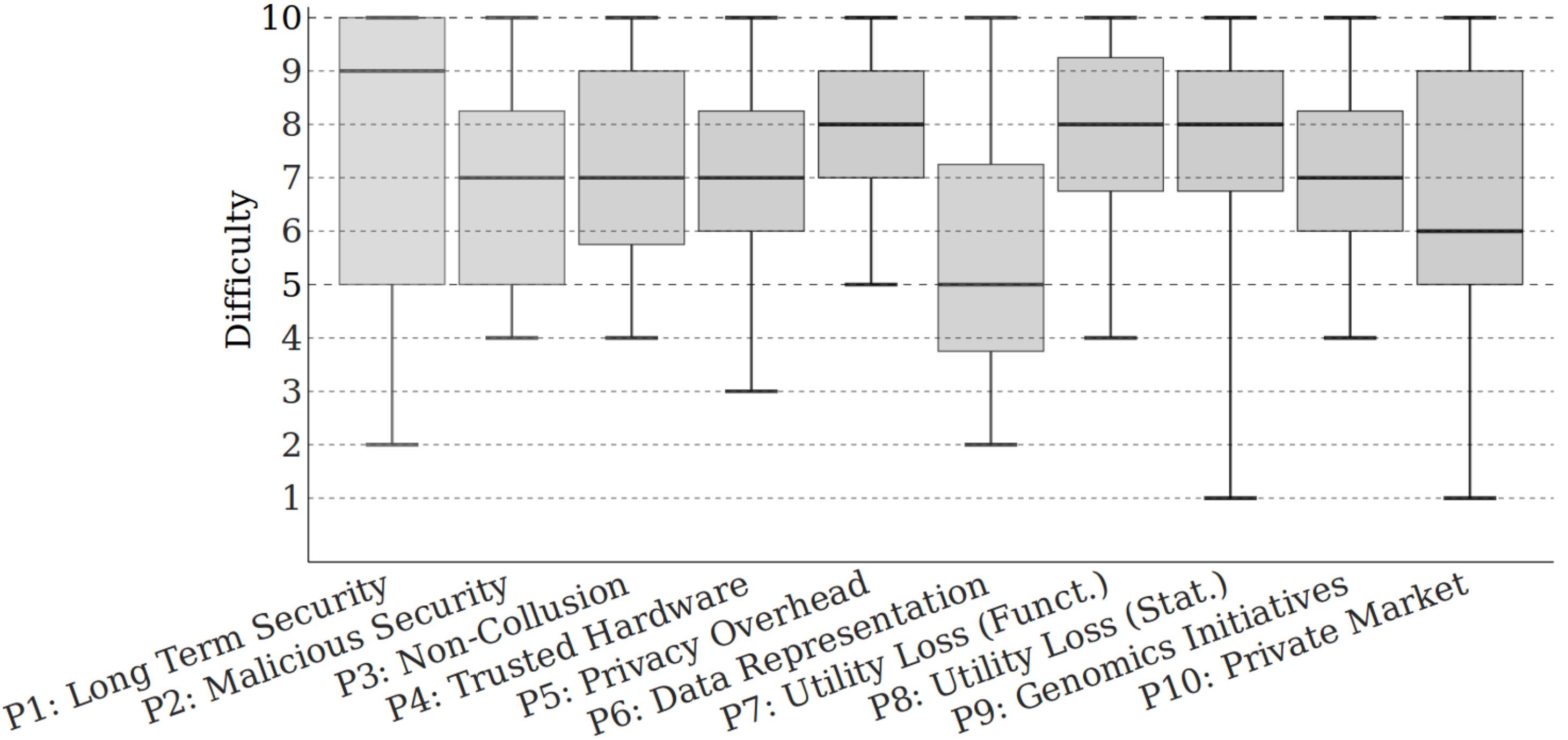}\label{fig:difficulty}}
\caption{Boxplots of answers to Q1 and Q2 of the online survey.}
\label{fig:boxplots}
\end{figure*}

The systematic analysis of research using PETs to protect genome privacy led to the identification of a list of ten technical challenges.
Aiming to validate and broaden the discussion around them, we sought the viewpoints of experts in the field with respect to their importance and difficulty.

\descr{Questionnaire.} We designed a short questionnaire, presenting participants with each of the ten challenges (P1--P10 in Section~\ref{sec:analysis}) and asking four questions for each:\smallskip
\begin{enumerate}
\item[~~~Q1.] How important is it to solve this problem?
\item[~~~Q2.] How difficult is it to solve this problem?
\item[~~~Q3.] What can be done to address this problem?
\item[~~~Q4.] How knowledgeable are you in this area?\smallskip
\end{enumerate}
\noindent For Q1-Q2, we used a ten-point Likert scale, with 1 corresponding to \enquote{not at all important/difficult} and 10 to \enquote{extremely important/difficult.} 
Q3 was a non-mandatory open-ended question, while Q4 provided three options: unfamiliar, knowledgeable, or expert.
The questionnaire took 10-15 minutes to complete.\footnote{A copy of the questionnaire is available from \url{http://mittos.me/wp-content/uploads/2018/08/questionnaire.pdf}.}

\descr{Participants.} To compile a list of genome privacy experts, we again used \url{GenomePrivacy.org}, exporting all the authors of the papers listed on the site (262 as of March 2018). 
We manually inspected each and removed those that appeared to have primarily biomedical or legal backgrounds, or only authored one paper, thus shortening the list to 92 names.
Then, we retrieved the email addresses of the authors from their websites 
and, in April 2018, we sent an email with a request to fill out the questionnaire.
After 30 days, we received answers from 21 experts.

The survey was designed to be {\em anonymous}, i.e., participants were provided with a link to a Google Form and were not asked to provide any personal information or identifiers.

\descr{Analysis.} Figures~\ref{fig:importance} and and~\ref{fig:difficulty} present boxplots of the answers to Q1 (importance) and Q2 (difficulty). 
For most questions, participants considered themselves experts or knowledgeable; only three identified as unfamiliar for P4:~Trusted hardware and P10:~(Relevance to) Private Market, two for P8:~Utility Loss (Statistical) and P9:~(Relevance to) Genomics Initiatives, and one for P3:~Non-Collusion, P5:~Privacy Overhead, and P6:~Data Representation.

Even though the questionnaire was administered with experts in the field, its limited sample and scope does not allow us to perform a quantitative analysis in terms of statistical hypothesis testing with respect to each problem.
However, the responses obtained for importance, along with the open-ended questions, do offer some interesting observations.\footnote{Numeric scores for difficulty are leveled around a median of 7, thus we do not discuss them due to space limitations.}
In fact, we found that participants seemed to consider most of the problems to be quite important overall. 
Looking at the median ($M$), we find that $M{\geq}7$ for eight out of ten problems, and $M{=}6$ for the other two: P2:~Malicious Security and P10:~Private Market.

\descr{Most Important Problems.} Three problems stood out as the most important, with $M{=}9$.
In particular, P1: Long-term Security received an average score of $\mu{=}8.23{\pm}2.02$, with $6$ out of the $7$ self-reported `experts' in this area giving it a $10$.
High scores were also obtained by P7:~Utility Loss (Functionality), with $\mu{=}8.85{\pm}1.16$, and P9:~Genomics Initiatives ($\mu{=}9.09{\pm}0.97$, the highest).
The importance of these problems was confirmed when normalizing scores by the average response of each participant, which let us reason about the ``relative'' importance; again, P1 and P7 stood out.

When asked about how to solve the problem of long-term security, the experts provided a wide range of ideas, thus confirming the need for future work that takes different approaches than the current ones.
For instance, some experts raised the availability of honey encryption~\mycite{huang2015genoguard}, which, however, is limited in scope and security (see Section~\ref{sec:long}).
Others proposed the use of gene manipulation techniques like CRISPR~\cite{ledford2015crispr}, blockchain-based solutions, and post-quantum cryptography. 
Specifically, one participant said: \enquote{use post-quantum or information-theoretic solutions. In theory, we know how to do this (just exchange primitives with PQ-secure ones), but one needs to check security and performance issues}.
Others focused on transparency of data access and management rather than confidentiality, stating, e.g., that \enquote{maybe cryptography is not the answer. Perhaps setting up an environment with different ways of controlling how the data is managed in order to provide more transparency.}
An expert proposed re-encrypting the data, 
while another suggested the use of large parameters: \enquote{One option is to use particularly large parameters, but this will degrade performance. Of course we can't know what improved attacks are coming up.}
Finally, a participant responded that this issue is not critical at the moment although it may be in the future, stating that \enquote{despite of the issues of impact to future generations, I do not consider this a critical factor in todays operations,}
while another simply responded that there is not much the community can do to address this issue. 
Whereas, when asked about utility loss with respect to functionality, there was some agreement that the community should start to carefully design specific solutions, rather than generic ones, and validate them with clinicians and researchers \textit{before} developing them.
For example, one participant mentioned: \enquote{we can develop fundamental and flexible evaluations primitives for various applications. Some solution seems to be the best for one case but there may be a better solution to other applications}, while another suggested to \enquote{work on practical use cases by collaborating with, e.g., medical researchers.} 
Further, an expert suggested to work on hybrid solutions: \enquote{The solution might have to be multi-layered with crypto, access control policy, etc. components}, while another focused on the users, suggesting to \enquote{educate humans to change their ways.}

\descr{Disagreements.} We also found that two problems, P8:~Utility (Statistical) and P10:~Private Market attracted very {\em diverse} scores, yielding a variance of, resp., $5.85$ and $6.80$.
This was confirmed by the participants' open-ended answers: four participants (self-identified as knowledgeable) rejected the use of differentially private mechanisms in clinical settings, while also providing low importance scores.
When asked how the community can address this challenge, some explained that, in certain cases, the utility loss can be manageable (e.g., \enquote{A comprehensive, experiment-driven analysis for feasible epsilon and delta values can pave the way towards more formally founded utility results for DP}), while another suggested that the utility loss is usually high because the proposed solutions are generic instead of specialized for a specific task.
A couple of participants did not recognize the issue of relevance to the private market as particularly important since they found privacy in this context to be fundamentally at odds with the DTC market, while others gave high scores for the very same reason.
\descr{Other Problems.} Among other observations, we found that P6:~Data Representation, although considered important ($M{=}9$), was also considered the easiest to address.
Specifically, it was suggested that solutions should be designed to better handle \textit{real} data, and that the community should focus on harmonizing existing schemes and improve interoperability.
Although this might be feasible, it once again highlights the need for truly interdisciplinary work.
By contrast, regarding the scarcity of solutions protecting against malicious adversaries (P2) or not relying on non-collusion assumptions (P3), we found that these were less important to the participants, both due to their confidence in advances in cryptography research but also because they felt these might be reasonable assumptions in the medical setting.

\descr{Take-Aways.} In summary, the 21 genome privacy experts who responded to our survey evaluated the ten challenges identified through our systematization, helping us put them in context and providing feedback on possible avenues to mitigate them. %
On the one hand, some of the issues are likely to be eventually mitigated via natural research progress (e.g., cryptographic tools geared to provide malicious security might become more efficient), optimization techniques (e.g., supporting specific functionalities in standardized settings), advances in new technologies (e.g., trusted hardware might reduce computation overheads), and/or inter-disciplinary collaboration (e.g., by improving interoperability).
On the other hand, however, we do not have a clear grasp as to how to tackle some of the other challenges that are inherently tied to the very nature of genome privacy.
For instance, the issue of long-term security cannot be effectively addressed using standard approaches; similarly, the utility loss stemming from hiding data might be hard to overcome with existing cryptographic and/or differential privacy techniques.

\reduceA\reduceA\reduceA\reduceA\reduceA\reduceA\reduceA
\section{Related Work}\label{sec:sok_vs_survey}
\reduceA\reduceA\reduceA

This section reviews prior work analyzing security and privacy challenges in the context of genomic data.

Erlich and Narayanan~\cite{erlich2014routes} analyze the different routes that can be used to breach and defend genome privacy.
They group plausible and possible attacks into three categories: completion, identity tracing, and attribute disclosure attacks, and discuss the extent to which mitigation techniques (i.e., access control, anonymization, and cryptography) can address them.
Ayday et al.~\cite{ayday2013chills} summarize the importance of progress in genomics along with the need for preserving the privacy of the users when dealing with their genomic data.

Shi et al.~\cite{shi2017overview} review genomic data sharing, the potential privacy risks, as well as regulatory and ethical challenges.
They also categorize tools for protecting privacy into controlled access, data perturbation (specifically in the context of differential privacy), and cryptographic solutions, providing an overview for each category.
Also, Wang et al.~\cite{wang2017genome} study the clinical, technical, and ethical sides of genome privacy in the United States, describing available solutions for the disclosure of results from genomic studies along record linkage, and the ethical and legal implications of genome privacy in conjunction to informed consent.

Naveed et al.~\cite{naveed2015privacy} present an overview of genome privacy work from a computer science perspective, reviewing known privacy threats and available solutions, and discuss some of the known challenges in this area---e.g., that genomic databases are often not under the control of the health system, or that privacy protection might affect utility. 
They also interview 61 experts in the biomedical field, finding that the majority of them acknowledge the importance of genome privacy research and risks from privacy intrusions.

Aziz et al.~\cite{aziz2017privacy} categorize genome privacy research in three groups: privacy-preserving data sharing, secure computation and data storage, as well as query or output privacy; they discuss solutions and open problems in those fields, aiming to help practitioners better understand use cases and limitations of available crypto primitives.
and compare their performance and security.
Finally, Akg{\"u}n et al.~\cite{akgun2015privacy} focus on bioinformatics research where private information disclosure is possible. 
For instance, they consider scenarios about querying genomic data and sequence alignment, while surveying available solutions for each.

\descr{Our SoK vs Prior Work.}
Overall, the current collection of surveys review and summarize available solutions. 
In combination, they complement our work and are useful in providing an overview of the state of the art, as well as the different avenues for privacy protection. Some also highlight challenges from non-technical perspectives, including ethics, law, and regulation.
However, they are mostly limited to reviews of already completed work by grouping and assessing available solutions without a structured methodology, and therefore, %
their analysis is usually restricted to the papers they survey and do not apply to the community in general.
Whereas, we define systematic criteria and sample relevant papers that are representative of a line of work;
this helps us provide a critical analysis of challenges and outlooks.
Specifically, we discuss several challenges which have not been identified in the past by previous surveys, including: the lack of long-term confidentiality protection, the over-reliance on non-collusion assumptions, the challenges presented by making strong data representations assumptions, as well as the lack of solutions applicable to current genomics initiatives.
Overall, our methodology can be reused in a straightforward manner in the future, e.g., to assess progress in the field, while, today, our analysis can be used  by the community as a guideline for future work.

\reduceA\reduceA\reduceA\reduceA\reduceA\reduceA\reduceA
\section{Conclusion}\label{sec:conclusion}
\reduceA\reduceA\reduceA

We introduced and applied a structured methodology to systematize knowledge around genome privacy, focusing on defensive mechanisms offered by privacy-enhancing technologies (PETs).
We selected a representative sample of the community's work and defined systematic criteria for evaluation.
We compiled a comparison table which guided a critical analysis of the body of work
and helped us identify ten technical challenges/open problems.
We also asked genome privacy experts to weigh in on how important and difficult they are to address and to provide ideas as to how to mitigate them. 
Overall, our analysis serves as a guideline for future work, while our methodology can be reused by other researchers to revisit new results and assess the progress in the field.

In short, we found that using PETs to protect genome privacy may be hindered by the obstacles related to the unique properties of the human genome. 
For example, the sensitivity of genome data does not degrade over time; as a consequence, one serious challenge stems from lack of long-term security protection, which is hard to address as available cryptographic tools are not suitable for this goal. 
We also observed that the overwhelming majority of proposed techniques, aiming to scale up to large genomic datasets, need to opt for weaker security guarantees or weaker models.
While it is not unreasonable to expect progress from the community with respect to underlying primitives, it is inherently hard to address the limitations in terms of utility and/or flexibility on the actual functionalities. 
When combined with assumptions made about the format and the representation of the data under analysis, this might pose major hurdles against real-life adoption.

These hurdles are further compounded by the interdependencies between some of the criteria %
and the categories %
discussed.
For instance, the use of cloud storage for genomic data implies the existence of a third party, and as such, the improvement in usability may be overshadowed by security limitations (Section~\ref{sec:sec-lim}). 
Furthermore, the solutions proposed in the Access \& Storage Control category may have a direct effect on every category as functionalities like secure storage and selective retrieval are crucial parts of any complete framework, further highlighting the importance of interoperability.   

Nevertheless, in its short lifespan, the genome privacy community has achieved admirable progress.
Indeed, a significant portion of research can already be used to enable genomics-related applications that are hard or impossible to support because of legal or policy restrictions.
For instance, genetic and health data cannot easily cross borders, which makes international collaborations very challenging. 
In this context, mechanisms provably guaranteeing privacy-friendly processing of genomic data may alleviate these restrictions and enable important research progress, and we hope to see more pilot studies along these lines in the near future.
In fact, some initiatives have started to provide encouraging results, e.g., a trial conducted at the CHUV hospital in Lausanne (Switzerland) to encrypt genetic markers of HIV-positive patients and let doctors perform tests in a privacy-preserving way~\cite{mclaren2016privacy}.

Furthermore, essential to the progress of the field are alternative and/or creative solutions to known problems. 
One such example is the work by Wan et al.~\cite{wan2017expanding}, who address the privacy-utility tradeoff in emerging genomic data sharing initiatives.  
To do so, they rely on a game-theoretic approach which accounts for the capabilities and the behavior of the adversary, so that the defender can choose the best strategy satisfying their privacy requirements without crippling utility.

Beyond the work in this paper, %
we also call for studying new genome manipulation techniques, such as CRISPR~\cite{ledford2015crispr}, and their potential impact on both safety (e.g., causing harm) and privacy (e.g., editing genomes to recover from data exposure or to hinder re-identification).
Finally, we plan to extend our analysis to privacy in broader biomedical contexts, such as considering users' health-related as well as genomic data.

\descr{Acknowledgments.} This research was partially supported by the European Union's Horizon 2020 Research and Innovation program under the Marie Sk\l{}odowska-Curie ``Privacy\&Us'' project (GA No. 675730), a Google Faculty Award on ``Enabling Progress in Genomic Research via Privacy-Preserving Data Sharing,'' and a grant from the National Institutes of Health (5RM1HG009034). We also thank the anonymous reviewers and our shepherd, Fabian Prasser, for their comments and feedback, as well as the experts who answered our survey.

{\small
\bibliographystyle{alphaabbr}
%\bibliography{bibliography}
\input{no_comments.bbl}}

\reduceA\reduceA\reduceA\reduceA\reduceA
\appendix %
\section{Genomics Primer}\label{primer}
\reduceA

\descr{Whole Genome Sequencing.} 
Whole Genome Sequencing (WGS) is the process of determining the complete DNA sequence of an organism.
In other words, it is used to digitize the genome of an individual into a series of letters corresponding to the various nucleotides (A, C, G, T).
WGS costs are now on the order of \$1,000~\cite{sequecing2017costs}. %

\descr{Whole Exome Sequencing.} 
Whole Exome Sequencing is the process of sequencing parts of DNA which are responsible for providing the instructions for making proteins. 

\descr{Genotyping.} 
Sequencing is not the only way to analyze the genome; in fact, in-vitro techniques such as genotyping are routinely used to look for known genetic differences using biomarkers.

\descr{SNPs, CNVs, STRs, RFLPs, Haplotypes, and SNVs.}
All members of the human population share around 99.5\% of the genome, with the remaining 0.5\% differing due to genetic variations. However, this 0.5\% variation does not always manifest in the same regions of the genome. As an artifact, storing how much or where differences occur could lead to uniqueness and identifiability concerns.
A Single Nucleotide Polymorphism (SNP) is a variation at a single position, occurring in $1\%$ or more of a population.
SNPs constitute the most commonly studied genetic feature today, as researchers use them to identify clinical conditions, predict the individuals' response to certain drugs, and their susceptibility to various diseases~\cite{welter2013nhgri}.
However, Copy Number Variations (CNVs)~\cite{stranger2007relative} and Short Tandem Repeats (STRs)~\cite{butler2007short} are also becoming increasingly more used.
Restriction Fragment Length Polymorphisms (RFLPs) refer to the difference between samples of homologous DNA molecules from differing locations of restriction enzyme sites, and are used to separate DNA into pieces and obtain information about the length of the subsequences.
A haplotype refers to a group of genes of an individual that was inherited from a single parent.
Finally, while a SNP refers to variation which is present to at least 1\% of a population, a Single Nuncleotide Variant (SNV), is a variation occurring in an individual, without limitations on frequency. 

\descr{GWAS.} 
Genome Wide Association Study is the process which compares the DNA of study participants to discover SNPs that occur more frequently in people carrying a particular disease.

\descr{SAM, BAM, FASTQ, and VCF.} 
Fully Sequenced Genomes (FGSs) are typically stored in either SAM, BAM, or FASTQ formats.
SAM (Sequence Alignment Map) is a text-based format, which may include additional information such as the reference sequence of the alignment or the mapping quality,
BAM is a binary format (in practice, the compressed and lossless version of SAM), while FASTQ is a text-based format which stores nucleotide sequences along with their corresponding quality scores.
VCF (Variant Call Format) is another text file format broadly used in bioinformatics for storing gene sequence variations.

\descr{Genome Operations.} 
A variety of operations can be performed on genomes. 
For the purpose of this paper, we describe three of the most common:

\begin{compactenum}

\item \emph{SNP weighted average.} There are several methods to compute the disease susceptibility of an individual, or other genetic testing related analysis that involve SNPs. A common method is based on weighted averaging, where certain weights are applied on SNPs to calculate the result of a test.

\item \emph{Edit distance.} An edit distance algorithm measures the dissimilarity of two strings. 
Specifically, the edit distance between two strings corresponds to the minimum number of edit operations (i.e., insertion, deletion, substitution) needed to transform one string into the other. 
In the context of genomics, usually researchers measure the edit distance between a patient and a reference genome.

\item {\em $\chi^2$ test.} A $\chi^2$ test is a statistical method to test hypotheses. 
Specifically, a $\chi^2$ test is used to determine whether there is a significant difference between the observed frequencies in a sample against the expected ones if the null hypothesis is true.
In the context of genomics, a $\chi^2$ test helps researchers determine whether certain hypotheses are true, e.g., determine whether two or more alleles are associated (linkage disequilibrium).
\end{compactenum}

\end{document}

%% file: no_comments.bbl
\newcommand{\etalchar}[1]{$^{#1}$}